\documentclass{mpe_report}

\usepackage{psfig,graphicx,epsfig}
\usepackage{color}
\usepackage{amsmath,amssymb,epic,eepic,array}

\begin{document}

\pagenumbering{arabic}
\setcounter{page}{76}

 \renewcommand{\FirstPageOfPaper }{ 76}\renewcommand{\LastPageOfPaper }{ 79}
\title{Detailed study of giant pulses from the millisecond pulsar B1937+21}
\author{V.I.~Kondratiev\inst{1,2}, M.V.~Popov\inst{1}, V.A.~Soglasnov\inst{1}, Y.Y.~Kovalev\inst{3,1,4}, N.~Bartel\inst{2}, \and F.~Ghigo\inst{3}}  
\institute{Astro Space Center of the Lebedev Physical Institute, Profsoyuznaya 84/32, Moscow, 117997 Russia
\and York University, Department of Physics and Astronomy, 4700 Keele Street, Toronto, Ontario M3J 1P3 Canada
\and National Radio Astronomy Observatory, P.O. Box 2, Green Bank, WV 24944, U.S.A.
\and Max-Planck-Institut f\"ur Radioastronomie, Auf dem H\"ugel 69, 53121 Bonn, Germany}

\authorrunning{V.~I.~Kondratiev et al.}
\maketitle

\begin{abstract}
The second fastest millisecond pulsar, B1937+21, is one of several
pulsars known to emit giant pulses (GPs).  GPs are characterized by
their huge energy, power-law cumulative energy distribution, and
particular longitudes of occurrence.
All these characteristics are different from those of regular
pulses. Here, we present a study of GPs from our observations of the
pulsar B1937+21 with the GBT at 2.1\,GHz in both left and right
circular polarization with a time resolution of 8\,ns. The Mark5 data
acquisition system was used for the first time in single-dish
observations with the GBT. This allowed us to obtain continuous and
uniform recording for 7.5~hours with a data rate of 512\,Mbps.  As a
result, more than 6\,000 GPs were found above a detection threshold of
200\,Jy.  We report on instantaneous spectra of GPs, as well as on a
comparison with scintillation spectra of regular emission, on the
distribution of GP energies, and on polarization properties of GPs.
\end{abstract}

\section{Introduction}
Giant pulses (GPs) are one of the most fascinating phenomena in pulsar radio
emission. They are manifested as separate huge pulses with intensities
hundred, thousand, and even million times larger than the intensity of
an average pulse.
Only 11 pulsars are known to emit such ``giant'' pulses. Among them,
only two -- the Crab pulsar and the original millisecond pulsar
B1937+21~-- have the strongest giant radio pulses ever
observed (e.g., Hankins et al. 2003; Popov et al. 2006a,b; Soglasnov et al. 2004).

Sallmen \& Backer (1995) presented the first analysis of GPs and
first noted that GPs occurred at the trailing edges of the main pulse
(MP) and the interpulse (IP) components. Then, extensive studies of
GPs from B1937+21 by ~\cite{cognard1996} and ~\cite{kink2000} were
done at Arecibo. \cite{cognard1996} detected 60 GPs at 430~MHz during
44 min of observations and reported up to about 100\% circular
polarization for some GPs. \cite{kink2000} carried out
non-simultaneous observations at 3 different frequencies: 430, 1\,420,
and 2\,380~MHz. Frequencies higher than 430~MHz allowed them to
determine the location of the occurrence of GPs. They found that GPs
occur in 10-$\mu$s windows delayed by $58~\mu$s and $65~\mu$s from the
maximum of the MP and IP components, respectively. A thorough
analysis of GPs was done in our previous study (\cite{soglasnov2004})
based on 39-min observations at 1\,650~MHz at Tidbinbilla with a high
time resolution of 31.25~ns. A number of 309~GPs were found, with the
strongest one having a peak flux density of 65~kJy. This peak flux
density together with a short duration of $<15$~ns corresponds to an
exceptionally high brightness temperature of $>5\times 10^{39}$~K. We also
determined the power-law index of the cumulative distribution of GP
energies to be $-1.4$.

Here, we present preliminary results of our study of GPs from
B1937+21 observed with the GBT with a high time resolution.

\section{Observations and data processing}

Observations of giant pulses from the millisecond pulsar B1937+21 were
done on June 7, 2005 between 02:30 and 10:40 UTC with the Robert
C. Byrd Green Bank Telescope (GBT) at a frequency of 2.1~GHz in both
left- and right-hand circular polarization (LCP, RCP) with a time
resolution of 8\,ns.  Four adjacent 16-MHz channels (2052--2116 MHz)
at each polarization were digitized simultaneously with 2-bit
sampling at the Nyquist rate. The Mark5A data acquisition system was
used for the first time in \emph{single-dish} observations with the
GBT (see GBT Commissioning Memo~236 by \cite{kovalev2005} for
information on how to use the GBT with a VLBA+Mark5A backend).  This
allowed us to obtain continuous and uniform recording for about
7.5~hours with a data rate of 512\,Mbps. The quasar 3C286, the radio
source 3C399.1 and the planetary nebula NGC~7027 were observed as
well, for flux density and polarization calibration.  The system
temperature in all 8 separate frequency channels was about 23~K. After
the observations the data were transferred from the Mark5 `8-pack'
disk modules to a Linux PC server at the observatory site. The
continuously recorded Mark5A data were split into individual pieces of
$10^9$ bytes. The size of the pieces was constrained by the storage
capacity.  Then, the split data were copied to external 1-TB disks and
shipped to home institutions for further processing.

The Mark5A data appear as if they were recorded to VLBA tapes. To get
the real 2-bit voltage signal from the Mark5A raw data, decoding
software was written\footnote{\tt
http://wiki.gb.nrao.edu/bin/view/Data/
HowToObserveReduceMark5AandS2Data}.  The obtained signal was corrected
for bit-statistics, i.e. for the fluctuation of instantaneous
rms values which results in a deviation of the signal levels away from
the levels for the optimum state of 2-bit
sampling (\cite{jenet1998}). Following \cite{hankins1971}, we then
dedispersed the data and corrected them for amplitude bandpass
irregularities.
Then we cut the signal with the topocentric pulsar period to search for GPs
and folded the signal to produce the average pulse profile.

It is known that GPs from B1937+21 occur in the trailing edges of the
MP and IP components of the average profile (\cite{soglasnov2004}).  We
chose a detection threshold of $17\sigma$ in every 16-MHz band and
searched for events in 60-$\mu$s windows at the trailing edges of
the regular MP and IP components. From all events stronger than our
threshold we selected ``true'' GPs if
a) they had intensities $>\!5\sigma$ in at least one other frequency
channel apart from the channel in which they were detected; b) their
intensity when reconstructed in the total 64-MHz passband was
$>\!5\sigma$; and c) they showed a characteristic scattering profile
for the pulses reconstructed in the 64-MHz passband.

\section{Results}

To date, 5.5 from 7.5 hours of pulsar data have been processed. We
detected 6\,334 GPs stronger than 205~Jy in the 16-MHz bands with the
strongest one having a peak flux density of 2~kJy (10~kJy in the total
64-MHz band).  The statistics for the GP occurrence are almost the
same both in the MP and IP regions as well as in LCP and RCP and in
the separate frequency channels.  The rate of GP occurrence is
about 20 GPs/min for the GPs exceeding the $17\sigma$-threshold in a
16-MHz band. The number of GPs with a high signal-to-noise ratio of
$>\!50$ in a 16-MHz band was found to be 177.

\begin{figure}
\centerline{\psfig{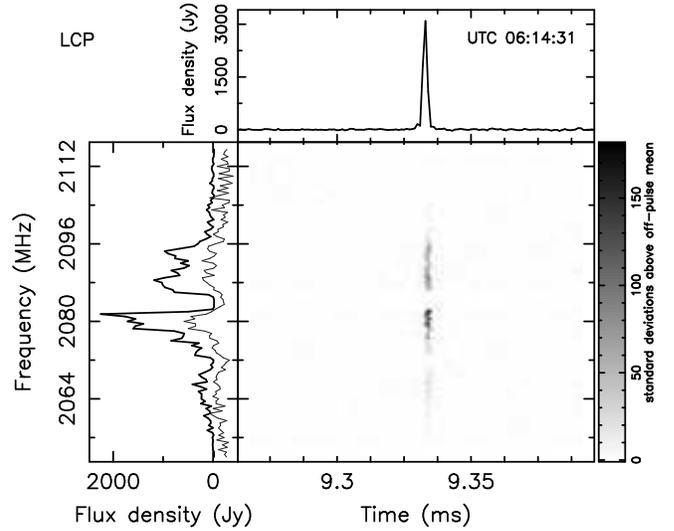} }
\caption{The instantaneous spectrum of one of the strongest GPs in LCP
shown with a spectral resolution of 0.5~MHz (128 channels).  The gray
scale shows the intensity of the GP in rms units. The plot at the top
represents the sum from all frequency channels giving the GP profile
with a time sampling interval of $1~\mu$s. The left plot (thick curve)
represents the spectrum of the GP averaged over three adjacent time
bins.  The thin curve in this plot shows the spectrum of the regular
emission over a 15-s time interval when the GP occurred.
\label{instspectrum}}
\end{figure}

\subsection{Interstellar scintillations}

\begin{figure}
\centerline{\psfig{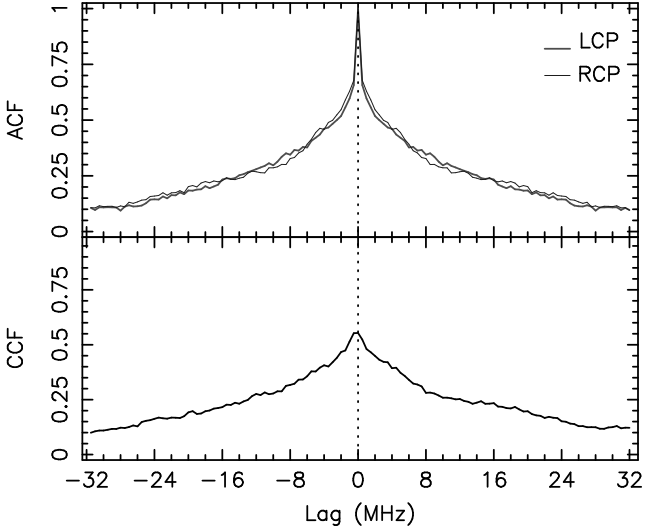}}
\caption{Average ACFs for LCP and RCP separately (top) and CCF between
the LCP and RCP data (bottom) for the spectra of 22 strong GPs.
The spectral resolution is 0.5~MHz.
\label{acfccf}}
\end{figure}

The influence of interstellar scattering on pulsar emission is much
smaller at 2.1~GHz than at lower frequencies, but is still
notable. The measured scattering time of individual GPs in our sample
was found to be about 40~ns, or 5 samples in the total 64-MHz
band. This measured scattering time gives a corresponding value of
decorrelation bandwidth of about 4~MHz. Computing instantaneous power
spectra of strong GPs, we found particular features in them (see
Fig.~\ref{instspectrum}).  To be precise, we computed the average auto
and cross-correlation functions (ACF, CCF) for the spectra in LCP
and RCP of 22 strong GPs (Fig.~\ref{acfccf}). We further measured a
frequency scale as the half-width at half the height between the
maximum and breakpoint in the CCF. It was found to be $\sim$4~MHz
which is about equal to the expected decorrelation bandwidth.

\begin{figure}
\centerline{\psfig{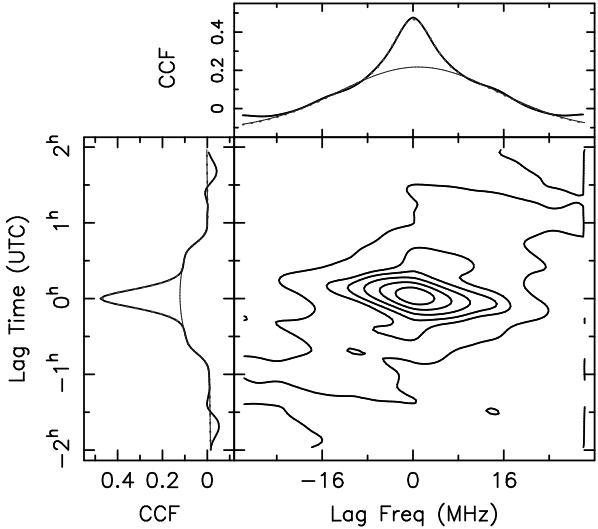} }
\caption{The 2-D CCF between dynamic spectra of regular emission in
LCP and RCP (center plot). The contours are plotted at 0, 18.4, 26,
36.8, 52, and 73.6~\% of the maximum. The plots at the top and left
show the central cut of the 2-D CCF in the frequency and time domain,
respectively (thick curves). The thin curves in these plots show the
fits used for the determination of the frequency and time scales.
\label{2dccf}}
\end{figure}

Are these features related to scattering or do they represent the
instrinsic spectrum of GPs?  The interstellar scintillation should
equally affect both regular and GP emission. Hence, we constructed the
dynamic spectra of regular emission in LCP and RCP that contain a
number of so-called ``scintles,'' spots of increased intensity in the
time-frequency domain. The 2-D CCF between these dynamic
spectra is shown in Fig.~\ref{2dccf}. Making cuts in frequency and
time, we measured two different scales.  In the frequency domain
a small scale was found to be $3.79\pm0.04$~MHz, and a large one
$16.5\pm 0.8$~MHz. The characteristic scintillation time is $10.4\pm
0.1$~min and $46.2\pm 0.4$~min, for the two different scales,
respectively. Only statistical standard errors are given.  The small
frequency scale of 3.8~MHz agrees well with the decorrelation
bandwidth found in the average CCF between the spectra of GPs. In the
left plot of Fig.~\ref{instspectrum}, the thin curve shows the
spectrum of regular emission in a 15-s time interval when a GP
occurred. Apart from a scaling factor, the spectrum of regular
emission is very similar to that of the giant pulse. Therefore we
conclude that the instantaneous spectrum of the GP reflects
scintillation in the ISM rather than the intrinsic spectrum of the GP.

\subsection{Energy distribution of GPs}

In contrast to regular pulse emission, the GP energies, E, are known
to obey power-law statistics with the rate of GP occurrence
$N_\mathrm{GP}\sim E^{\alpha}$ (Kinkhabwala \& Thorsett 2000) and with a low-energy
but no high-energy cut-off (\cite{soglasnov2004}). To study the
statistics of low-energy GPs and search for the low-energy threshold
we have constructed the cumulative distribution of GPs for every
frequency channel and polarization.  All determined indexes are
summarized in Table~\ref{energydistr}.  It was found that power-law
indexes have a large jitter around the mean value of $-2.2$. This
seems to be caused by scintillation.  The mean value of $-2.2$ differs
significantly from the value of $-1.4$ we found in our previous
study (\cite{soglasnov2004}).  However, we also found that the
cumulative distributions for half of the channels may be approximated
by power-law functions with two different slopes, $\alpha_1$ and
$\alpha_2$ (see Table~\ref{energydistr}).
The values of $\alpha_2$ are close to $-1.4$ from our previous study
(\cite{soglasnov2004}) which had less sensitivity. Hence, moving
downwards toward lower energies the cumulative energy distribution of
GPs gets steeper and also agrees better with the power-law index
of $-2.3$ obtained for the Crab pulsar (\cite{lundgren1995}).

\begin{table}
      \caption{Power-law indexes of the cumulative distribution of
      GPs. The first column gives the polarization, the second column
      lists the corresponding frequency channels, column 3 lists
      values of power-law indexes $\alpha_1$ in the low-energy region,
      and column 4 gives the values $\alpha_2$ of the power-law
      indexes in the high-energy region, if available. The last row
      presents the mean value of $\alpha_1$ among all frequency
      channels. Errors are statistical only.}

         \label{energydistr}
      \[
         \begin{tabular}{lccc}
            \hline
            \noalign{\smallskip}
            {\rm Pol} & {\rm Channel}    &  ${\alpha}_1$ & ${\alpha}_2$ \\
                      & {\rm (MHz)}      &  & \\
            \noalign{\smallskip}
            \hline
            \noalign{\smallskip}
	     LCP & 2100~--~2116 & $-2.03\pm0.02$ & $-1.1\phantom{8}\pm0.1\phantom{4}$ \\
	         & 2084~--~2100 & $-3.73\pm0.06$ & $-1.4\phantom{8}\pm0.3\phantom{4}$ \\
		 & 2068~--~2084 & $-2.48\pm0.04$ & \\
		 & 2052~--~2068 & $-2.43\pm0.02$ & $-1.38\pm0.01$ \\
            \noalign{\smallskip}
	    \hline
            \noalign{\smallskip}
	     RCP & 2100~--~2116 & $-2.63\pm0.05$ & $-1.42\pm0.04$ \\
	         & 2084~--~2100 & $-2.22\pm0.03$ & \\
		 & 2068~--~2084 & $-3.15\pm0.04$ & \\
		 & 2052~--~2068 & $-2.25\pm0.02$ & \\
            \noalign{\smallskip}
	    \hline
            \noalign{\smallskip}
	    \multicolumn{2}{l}{Mean} & $-2.2$ & \\
            \noalign{\smallskip}
             \hline
         \end{tabular}
      \]
   \end{table}

\subsection{Polarization of GPs}

\begin{figure}
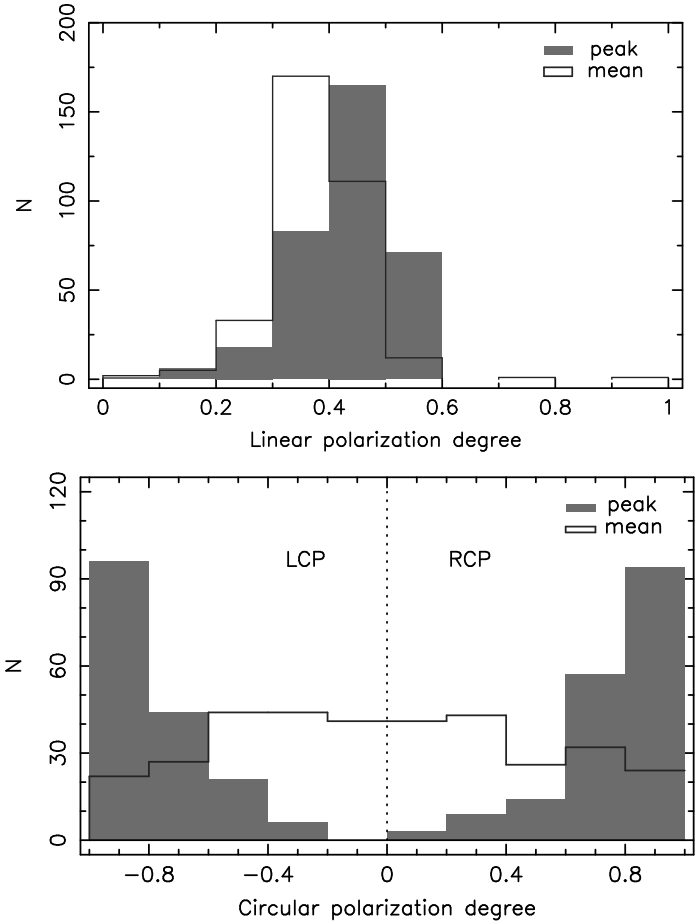

\centerline{\psfig{file=lin.ps,width=6.0cm,angle=-90,clip=} }
\vskip 2mm
\centerline{\psfig{file=circ.ps,width=6.0cm,angle=-90,clip=} }
\caption{Histograms that represent the number of GPs versus fractional
linear (top) and circular (bottom) polarization. The total number of
selected GPs with a signal-to-noise ratio in a 16-MHz band of $>\!25$
is 343. Histograms in gray represent values for the peaks of the GPs,
while open histograms represent average values for the whole GP.
\label{pol}}
\end{figure}

We selected 343 strong GPs with a signal-to-noise ratio in a
16-MHz band $>\!25$, sufficiently high to study the polarization
properties of GPs. In Fig.~\ref{pol} we present the distribution of
GPs versus linear and circular fractional polarization.  It can be
seen that the majority of the GPs ($>\!55$\%) have circularly
polarized peaks with fractional polarization $>\!0.8$ (either left, or
right). Only a few of the GPs (15\%) have fractional circular
polarization $<0.6$.  The fractional linear polarization of GPs is
also very high. Out of the 343 GPs, 165 (48\%) have fractional linear
polarization of 0.4--0.5. It should be mentioned that the phase of the
peak in circular and linear polarization profiles within the same GP
could be different. Hence, the same GP can reveal both strong circular
and strong linear polarization.  The number of GPs with both large
circular ($>\!0.8$) and large linear ($>\!0.4$) fractional
polarization is 130, or almost 38\%. Thus, GPs are very strongly,
both, circularly and linearly polarized events.

\section{Conclusions}

The Mark5 recording system provides an excellent possibility to study
properties of GPs with high time resolution at single-dish radio
telescopes.  Preliminary data reduction of the observations of the
millisecond pulsar B1937+21 made with the GBT at a frequency of
2.1~GHz confirms the very short duration of GPs of $<8$~ns.
The instantaneous spectrum of a GP was found to correspond well to the
diffraction spectrum of regular emission.
In half of cases
the cumulative distribution of GP energies may be represented by a
power-law piecewise function with two slopes and indexes of $-2.2$ and
$-1.4$. For the first time the polarization properties of the GPs of
pulsar B1937+21 were analyzed.  Almost all strong GPs are highly
circularly polarized, with $>\!55$\% of GPs having
fractional circular polarization $>\!0.8$.

The work reported in these proceedings is in progress. The complete 
analysis will be published later. 

\begin{acknowledgements}
VIK is grateful to Walter Brisken for his help with decoding the
Mark5A data format.  The Robert C. Byrd Green Bank Telescope (GBT) is
operated by the National Radio Astronomy Observatory which is a
facility of the U.S. National Science Foundation operated under
cooperative agreement by Associated Universities, Inc.  VIK was a
postdoctoral fellow at York University at the beginning of this
project.  This project was done while YYK was a Jansky Fellow of the
National Radio Astronomy Observatory and a research fellow of the
Alexander von Humboldt Foundation.  This project was supported in part
by grants from the Canadian NSERC, the Russian Foundation for Basic
Research (project number 04-02-16384) and the Presidium of the Russian
Academy of Sciences ``Origin and evolution of stars and galaxies.''
VIK acknowledges support from the Science and Innovation Federal
Agency (grant of the President of Russian Federation
MK-4032.2005.2). The participation of VIK and VAS in the 363rd Heraeus
Seminar on Neutron Stars and Pulsars ``About 40 years after the
discovery'' was supported in part by the WE-Heraeus foundation.
\end{acknowledgements}



    \clearpage

\end{document}